\documentclass[a4paper]{article}
\usepackage{graphics}
\usepackage{subfigure}

\begin{document}

\title{\noindent {\bf {\Large 
TIGHT-BINDING LINEAR SCALING METHOD APPLICATIONS TO SILICON SURFACES
}}}

\author{
{\bf 
Abduxukur~Abdurixit $^{1}$, Alexis~Baratoff $^{1}$ and
Giulia~Galli $^{2}$
} \\  \\ 
\small \it 
$^{1}$Institute of Physics, University of Basel, Klingelbergstr. 82,
4056 Basel, Switzerland\\
\small \it
$^{2}$Lawrence Livermore National Laboratory,
P.O.Box 808, L-415, Livermore, CA 94551, USA
}

\date{}

\maketitle



\baselineskip 11.pt

\begin{abstract}

The past years have witnessed impressive advances in electronic structure
calculation, especially in the complexity and size of the systems studied,
as well as in computation time. Linear scaling methods based on empirical
tight-binding Hamiltonians which can describe chemical bonding, and have a
computational time proportional to the number of atoms N in the system,
are of particular interest for simulations in material science. By
contrast conventional diagonalization schemes scale as $N^{3}$. In
combination with judiciously fitted parameters and an implementation
suited for MD, it is possible to apply such \emph{O(N)} methods to
structural, electronic and dynamical properties of large systems which
include up to 1000 atoms on a workstation.

Following a brief review of a tight-binding based linear-scaling method
based on a local orbital formulation and of parametrizations appropriate
for covalently bonded systems, we present recent test calculations on
\emph{Si(111)-5 $\times$}5 and \emph{Si(001)-c(4$\times$}2) reconstructed
surfaces in this framework, and compare our results with previous
tight-binding and \emph{ab-initio} calculations.\\

\end{abstract}

\pagestyle{myheadings}
\markboth{Tight-Binding Linear-Scaling Method and Applications to Silicon
Surfaces\\ 
\hskip 1.2cm 
{\small Abduxukur Abdurixit {\it et al}}}
{Tight-Binding Linear-Scaling Method and Applications to Silicon
Surfaces\\ 
\hskip 1.2cm 
{\small Abduxukur Abdurixit {\it et al}}}

\section{Introduction: \emph{Ab Initio} vs. Semiempirical,
$O(N^{3})$ vs. $O(N)$ Methods}

In the past two decades, the percentage of theoretical investigations of
materials based on atomistic computer simulations has steadily increased.
Those using either an \emph{ab-initio} or a tight-binding independent
electron description of interactions are able to account for chemical bond
formation and breaking, and are particularly worthy of attention, as
reflected in the 1998 Nobel prize for chemistry. One shortcoming of the
usual implementation of such methods is that the number of operations
scales as the third power of the number of atoms N, i.e., their
computational cost is $O(N^{3})$. Currently this limits the number of
atoms in the system that can be treated even using very powerful
computers. \emph{Ab-initio} methods certainly give more precise results
than semiempirical tight-binding methods, but besides being based on more
complicated Hamiltonians, they require much more extensive basis sets to
expand the wave functions of electrons. Empirical tight-binding methods
provide a useful compromise between classical empirical potential
approaches and \emph{ab-initio} methods, because they retain a quantum
mechanical description of the electrons, ultimately responsible for
chemical bonding, but the Hamiltonian is parametrized and the
wavefunctions expanded in a minimal set of atomic orbitals. As a
consequence, the number of atoms which can be handled with even the
simplest \emph{ab-initio} method, the Local Density
Approximation\cite{kohn-sham} commonly used in materials science is one to
two orders of magnitude less than with tight-binding methods within the
same computational constraints . Whereas chemists are interested in
studying large, complex molecules, materials scientists are concerned with
the properties of clusters, solids with specific defects or disorder,
surfaces, interfaces, artificial structures and their interactions.
Reliable computations of properties require simulations on large enough
finite systems, e.g. enclosed in a ``box'' with periodic boundary
conditions applied.

In order to increase the number of atoms in the system and to study
dynamical process or finite-temperature properties obtained from time
averages, efforts are constantly made to reduce computational cost. For
this reason, many kinds of linear scaling methods have been introduced,
tested and compared . The interested reader is referred to recent
reviews\cite{galli1996,bowler1997}. Linear scaling or \emph{O(N)} means
that the computation time is proportional to the number N of atoms in the
system, just like in classical simulations with finite-range interaction
potentials\cite{AllenTildesley}. In this contribution we concentrate on a
particular orbital-based linear scaling method which, in conjunction with
a tight-binding Hamiltonian, has been successfully applied to \emph{C} and
\emph{Si} systems\cite{mauri1994,kim1995}.

This contribution is organized in following way: First we review the
approximations and the method, then we present results on
$Si(111)-5\times5$ and $Si(001)-c(4\times 2)$ reconstructed surfaces which
serve to validate the method for future applications. At the end we
compare our results with previous LDA and tight-binding computations and
summarize our conclusions.

\section{The Tight-Binding Linear Scaling Method}

\subsection{Semiempirical Tight-Binding Approximation}

The empirical tight-binding (TB) approximation allows the quantum
mechanical nature of covalent bonding to enter the interaction Hamiltonian
in a natural way rather than through additional \emph{ad hoc} angular
terms in a classical potential.

In TB models the total energy of the system is expressed as
\begin{equation}
\label{tot}
E=E_{BS}+\sum _{LL'}\phi\left( \left|
\mathbf{R}_{L}-\mathbf{R}_{L'}\right| 
\right), 
\end{equation}
where $\phi$ is a repulsive two-body potential which includes the ion-ion
repulsion and the electron-electron interactions which are double counted
in the electronic "band-structure" term $E_{BS}$. This term describes
chemical bonding, it can be written as,
\begin{equation}
\label{band}
E_{BS}=\sum _{i}f_{i}\epsilon_{i}=\sum _{i}f_{i}<\psi _{i}\left|
\hat{H}_{TB}\right| \psi _{i}>,
\end{equation}
where $\hat{H}_{TB}$ is the TB Hamiltonian, $\epsilon_i$ and
$\{\psi_{i}\}$ are its eigenvalues and eigenstates, and $f_i$ is their
occupancy. The total number of valence electrons is from now on denoted as
N and the total number of single particle states is then N/2 if we assume
double occupancy, i.e., spin degeneracy for each state($f_{i}=2$). The
occupied eigenstates can in principle be determined by diagonalization,
but in our work diagonalization is only used at the very end of each
computation to check its accuracy. The off-diagonal elements of
$\hat{H}_{TB}$ are described by invariant two-center matrix elements,
$V_{ss\sigma },\, V_{sp\sigma },\, V_{pp\sigma } $ and $ V_{pp\pi }$,
between the set of $sp^3$ atomic orbitals(assumed orthonormal). By
adjusting their values at the interatomic distance $r_0$ = 2.35\AA ~in the
equilibrium diamond-like structure, as well as diagonal elements $E_s$,
$E_p$, a good fit to the position and dispersion of the occupied valence
bands of Si can be obtained~\cite{chadi1979-jvst}.

 In order to study properties of covalently bonded systems with defects or
free surfaces tight-binding model must be transferable to different 
physically relevant environments. An important advance was due to Goodwin,
Skinner and Pettifor\cite{gsp} who showed that it is possible to set up a
TB model which accurately describes the energy-versus-volume behaviour of
Si in crystalline phases with different atomic coordination, and
reproduces the structure of small Si clusters.

We therefore adopt the functional form suggested by GSP\cite{gsp} for the
distance dependence of the two-center matrix elements and of the two-body
potential:
\begin{equation}
\label{scalingfunction}
V_{\alpha}(r)=V_{\alpha}(r_{0})\left( \frac{r_{0}}{r}\right)
^{n}exp\left\{ n\left[ -\left( \frac{r}{r_{c}}\right) ^{n_{c}}
+\left( \frac{r_{0}}{r_{c}}\right) ^{n_{c}}\right] \right\} 
\end{equation}
\[(\alpha \, :\, ss\sigma ,\, sp\sigma ,\, pp\sigma ,\, pp\pi )\]
\begin{equation}
\label{repulsivepotential}
\phi (r)=\phi _{0}\left( \frac{r_{_{0}}}{r}\right) ^{m}
exp\left\{ m\left[ -\left( \frac{r}{d_{_{c}}}\right) ^{m_{_{c}}}
+\left( \frac{r_{_{0}}}{d_{_{c}}}\right) ^{m_{_{c}}}\right] \right\} 
\end{equation}
where $r$ denotes the interatomic separations, and $n<<n_c$, $m<<m_c$,
$r_c \approx d_c > r_0$ and $\phi_0$ are parameters which are
determined by fitting judiciously chosen properties. The resulting high
values of $n_c$, $m_c$ ensure a rapid decay of $V_\alpha(r)$, $\phi(r)$
beyond $r_c$, $d_c$. In molecular dynamics(MD) simulations where a finite
range of $r$ is explored, the quantities in
Eqs.(\ref{scalingfunction}-\ref{repulsivepotential}) are further
multiplied by a smoothed step function which switches from 1 to 0 in a
narrow interval about a cutoff radius $r_m > r_c, d_c$.

As explained below, similar couplings between Si and H atoms are required
in some of our computations. Because hydrogen has a single occupied {\it
s}-state, only $ss\sigma$ and $sp\sigma$ matrix elements must be
parametrized. $H-H$ couplings are negligible at the separations considered
here.

\subsection{Tight-Binding Parameters}

Within the above framework, improved sets of parameters were subsequently
developed for Si-Si and Si-H interactions. In this contribution, we
used the two different sets of parameters described in detail by Kwon
\emph{et al}\cite{kwon1994} and Bowler \emph{et al}\cite{bowler1998}.

The Si-Si parametrization of Kwon \emph{et al}\cite{kwon1994} is rather
complicated; thus $r_c$'s, $n_c$'s depend on $\alpha$ and the repulsive
contribution is represented by a nonlinear function of the second term in
Eq.(\ref{tot}). The parameters are fitted to many properties of \emph{Si}
in the diamond structure and to computed(LDA) cohesive energies vs.
density of four different structures. The resulting properties of liquid
Si and of small Si clusters are in remarkable agreement with experiments
and with {\it ab-initio} computations. On the other hand, this
parametrization uses a cut-off which is beyond second nearest neighbours;
this implies significantly larger computing times. The revised GSP
parametrization of Bowler \emph{et al} \cite{bowler1998} is less
complicated than the previous one; \emph{Si-Si} parameters were fitted to
fewer properties in the diamond and $ \beta $-tin structures, \emph{Si-H}
parameters to properties of $SiH_4$. Furthermore, the cut-off radius $r_c$
can be chosen between nearest and next-nearest-neighbours; thus this
parametrization is very well suited for linear scaling computations. It
has in fact been successfully applied to defects and hydrogen diffusion on
Si(001)~\cite{bowler1998}.

\subsection{Orbital Based Linear Scaling Energy Minimization without
Orthogonalization Constraints}

Traditional electronic structure methods solve the Schr\"odinger equation
by expanding one-electron wavefunctions in a fixed basis set (plane waves,
atomic orbitals or combinations thereof) and by diagonalizing the
resulting secular equation for the expansion coefficients. In spite of
significant progress achieved by applying efficient diagonalization
algorithms, the required computing time is proportional to $NP^{2}$,
\emph{P} being the number of basis functions. Because $P\propto N$, the
computational cost is $O(N^3)$, the scaling factor depending on the
method, being small in the case of empirical tight-binding. Nevertheless,
the diagonalization of $\hat{H}$ for each atomic geometry or at each step
of MD simulation limits the number of atoms that can be currently studied
in conventional TBMD computations to about 100 using a workstation and
around 1000 using a supercomputer. The recently developed linear scaling
methods compute the total energy by minimizing a functional expressed in
terms of localized orbitals in real space. Although typical eigenstates in
a condensed system extend throughout most of it, a unitary transformation
yields linear combination of the former which are localized about
particular sites. On the basis of exact model calculations~\cite{kohn},
these so-called Wannier orbitals are believed to decay exponentially in
systems with a finite energy gap between occupied and empty
states~\cite{marzari}. This applies in particular to the finite systems on
which computations are carried out, although the effective gap can be
small and corresponding decay slow if the corresponding real system
is metallic.

The key feature of $O(N)$ methods is that the total energy and the forces
acting on individual atoms are evaluated \emph{without computing the
eigenvalues and eigenstates of $\hat{H}$}. This is accomplished by
dividing the full system into finite subsystems and by defining {\it
localized orbitals} ~\{$\phi$\} which are forced to vanish outside each 
subsystem\cite{galli1992}. These \emph{localized regions} (LR) are the
electronic equivalents of the \emph{linked cells} which ensure \emph{O(N)}
scaling in classical simulation\cite{AllenTildesley}. Intuition and
experience suggest that the minimum size of each localization region
depends on physical and chemical properties of the constituents, and not
on the size of the system whole. The size of the localized regions (which
must exceed the range of $\hat{H}$) is always the factor which limits the
accuracy of an $O(N)$ calculation.

Another key ingredient to achieve $O(N)$ scaling is the definition of an
appropriate energy functional whose minimization requires neither explicit
orthogonalization of the auxiliary electronic orbitals, nor the inversion
of their overlap matrix $\mathbf{S}$. This functional is in general
different from that defined in Eq.(\ref{tot}), but must have the same
global minimum in the limit of infinite localization regions. Otherwise it
yields an upperbound which in practice, must be close to the minimum
energy even for relatively small LR's. This can in fact be
achieved~\cite{galli1996,bowler1997,kim1995}. Various \emph{O(N)} methods
are based on different functionals which, however, share this remarkable
property.

One convenient energy functional which satisfies these requirements
is~\cite{kim1995}:
\begin{equation}
\label{3}
E_{GBS}\left[ \left\{ \phi \right\} ,\mu ,M\right] 
=2\sum ^{M}_{ij=1}(2\delta _{ij}-S_{ij})<\phi _{j}\left| \hat{H}
-\mu \right| \phi _{i}>+\mu N
\end{equation}
The matrix $(2I-S)$ is the first order truncated series expansion of the
inverse overlap matrix $S^{-1}$, where $S_{ij}=<\phi _{i}|\phi _{j}>$. The
functional defined in Eq.(\ref{3}) depends on the number M localized
orbitals (LO), and on a global variable $\mu$ which determines the highest
filled state and hence the total electronic charge in the system at the
minimum. Taking $M > N/2$ helps avoid unphysical solutions~\cite{kim1995},
depending on the initial choice of the orbitals, which can otherwise be
obtained.

In order to find the electronic ground state energy for a given spatial
configuration of the atoms, the functional is minimized with respect to
the LO's. Normally each LR is centered at an atomic site \emph{I} and
encompasses all neighbours connected by n bonds; it is then denoted by
nLR. Then LO $\phi _i$ centered at atomic site \emph{I} can be expressed
as
\begin{equation}
\label{4}
\phi _{i}=\sum ^{n_{b}}_{J\in \{LR_{I}\}}\sum _{l}C_{Jl}^{i}\alpha _{Jl}
\end{equation}
where $\alpha _{Jl}$'s are the atomic basis functions of atom \emph{J},
the index \emph{l} runs over orbital components (e.g.  $s,\, p_{x},\,
p_{y},\, p_{z}$ for carbon or silicon and s for H), $\{LR_{I}\}$ indicates
the set of atoms within the localization region centered at site \emph{I},
and $n_b$ is the corresponding number of basis functions.

The functional is efficiently minimized using a conjugate gradient
(CG) algorithm\cite{NumericalRecipes}. The required derivatives
\begin{equation}
\label{5}
\frac{\partial E_{GBS}}{\partial \phi _{i}}
=4\sum _{j}^{M}\left[ \left| (H-\mu )\phi _{i}>(2\delta _{ij}-S_{ij})
-\left| \phi _{j}><\phi _{j}\left| (H-\mu )\right| \phi _{i}>\right. \right. 
\right] 
\end{equation}
are evaluated at each iteration step. Using Eq.(\ref{4}) the matrix
elements can be expressed in terms of the Slater-Koster energy integrals
$<\alpha_{Jl}|\hat H_{TB}|\alpha_{J^{'}l^{'}}>$~\cite{slater-koster}; 
those energy integrals can be further expressed in terms of direction
cosines and of the invariant two-center matrix elements defined by
Eq.(\ref{scalingfunction})(see Appendix B). Because each derivative needs
to be evaluated only in the localization region $\{LR\}_ I$, the required
number of operations scales linearly with the number of atoms in the
system. The global variable $\mu$ is initially chosen well above the
estimated Fermi energy of the system, then iteratively adjusted until the
total charge of the system becomes equal to the charge consistent with
global neutrality when convergence is achieved, i.e., the ground state
energy corresponding to the assumed LRs is attained, and $\mu$ is the
chemical potential (Fermi energy) of the electrons\cite{kim1995}.

\subsection{Total Energy Minimization with Respect to Atomic Positions and
Electronic Readjustment}

In this work we consider only known metastable structures of silicon
surfaces in order to test the performance and accuracy of the method. To
determine such structures, atoms in several surface layers are allowed to
move under the influence of the forces
\begin{equation}
\label{7}
F_{I}=-\partial E/\partial R_{I}
\end{equation}
until all their components become smaller than a preset tolerance (0.01
\emph{eV/\AA}) and the energy \emph{E} reaches a minimum. Instead of using
a standard minimization procedure, this is achieved by introducing a
damping term in the standard Verlet algorithm~\cite{AllenTildesley}. This
term is adjusted so that the resulting motion is almost critically damped,
so that the fastest possible relaxation is achieved.

Eq.(\ref{7}) is physically meaningful only if the electrons are in their
ground state (Born-Oppenheimer approximation). Therefore, molecular
dynamics can be started only after electronic convergence with respect to
the initial atomic configuration has been achieved as described in section
2.3. This rather tedious procedure is necessary at the start of a
calculation. The required computing time depends on the initial choice of
the LOs and $\mu$. To be safe, we start with random coefficients in
Eq.(\ref{4}) and a high $\mu$. To ensure stable MD integration, the time
step must be small compared to a typical optical vibration period. The
corresponding atomic displacements are then small and typically do not
strongly perturb the LO coefficients, except if same atom(s) move out or
into certain LRs. Therefore, after each atom move enough electronic
iterations must be performed in order to reach the slightly modified
ground state (within a relative tolerance $ < 10^{-4}$).
Fortunately, only small adjustments of $ \mu $ are necessary once global
charge neutrality has been established; they can be performed
automatically. The number of electronic steps depends on the system, the
narrower its energy gap at the Fermi level(always existing in a finite
system), the more electronic steps are needed. Useful \emph{O(N)}
performance is achieved if this number of electronic iterations is
independent of N and if the necessary computing time is less than that
required for diagonalization.

\subsection{Local charge neutrality}

If significant atomic rearrangements occur in the course of a molecular
dynamics simulation, it is sometimes difficult to avoid unphysical charge
transfer between neighboring atoms or layers. To reduce such effects,
which slow down convergence and sometimes lead to unphysical solutions,
local charge neutrality can be approximately imposed by adding an extra
term $H_{U}=\frac{1}{2}U\sum _I(q_{I}-q_{I}^{0})^2$, where
$q_{I}^{0}=4.0$ for \emph{Si} and 1.0 for {\it H} atoms, and the charge
around site \emph{I} is expressed as
\begin{equation}
\label{8}
q_{I}=2\sum _{ij}(2\delta _{ij}-S_{ij})<\phi _{i}\left| \mathbf{R}_{I}>
<\mathbf{R}_{I}\right| \phi _{j}>
\end{equation}
where $<\phi _{i}|\mathbf{R}_{I}>$ indicates the projection of the
localized orbital $\phi _i$ on the localized region around site
\emph{I}~\cite{kim1995}. Such a term is obtained by making a local mean
field approximation to the Hubbard Hamiltonian and assuming no
spin-polarized solutions\cite{alerhand1987}. The strength \emph{U} of the
Hubbard-like term has been estimated and tabulated by
Harrison\cite{Harrison} who also discussed its reduction by dielectric
polarization. In covalent systems such Hubbard-like contributions must
essentially vanish once convergence is achieved, for atoms which have a
bulk-like environment.

\section{Applications }

In this section we present our recent test calculations with the method
described above which has hitherto been mainly applied to carbon systems.
Our ultimate goal is to simulate atomic force microscopy(AFM) and
manipulation with a Si tip. We present and discuss our results on
\emph{Si(111)} and \emph{Si(001)} surface reconstructions which have
previously been studied by tight-binding and \emph{ab-initio} methods, and
exhibit characteristic features due to rebonding of surface atoms. Such
reconstructions reduce the density of energetically unfavourable "dangling
bonds" on surface atoms, but induce distortions from the tetrahedral
bonding pattern in the bulk. Optimum surface structures represent a
delicate balance between different effects and can not usually be
guessed by chemical and physical intuition. 

These reconstructed surfaces are therefore well-suited for validating the
method and are also interesting candidates for controlled atom 
manipulation experiments. Surface properties are extracted from
computations on slabs with a finite number of layers. This number must be
large enough to suppress artificial coupling between the free slab
surfaces which can arise owing to the overlap of surface states and/or to
strain fields induced by atomic rearrangements in surface layers. In order
to approximate the effect of the crystalline substrate, atoms in the two
centeral layers are held in their bulk positions. Alternatively, the
bottom two layers are fixed, and all exposed dangling bonds are passivated
with hydrogen atoms so as to preserve tetrahedral coordination. Similar
accuracy is expected with half the number of free layers, compared to the
former, symmetric slab. Having this in mind, we have investigated the
influence of factors which, if chosen properly, should have little effect
on physical meaningful results. This includes the number of free and fixed
(bulk-like) layers in the slabs used, the influence of H-passivation at
the bottom, the shape and lateral dimensions of the computational
supercell, the size of the localization regions, the tight-binding
parametrization, starting from configurations which had the periodicity
and some of the rebonding characteristics of desired reconstructions. All
this required many computations which were performed on not too large
systems in order to economize computing time. Thus the emphasis here is on
validation rather than on achieving \emph{O(N)} performance, although this
was demonstrated in the larger systems containing a few hundred atoms.

\subsection{Si(111)-5x5 Reconstruction  }

\begin{figure}[!ht]
{\centering \resizebox*{10cm}{7cm}{\includegraphics{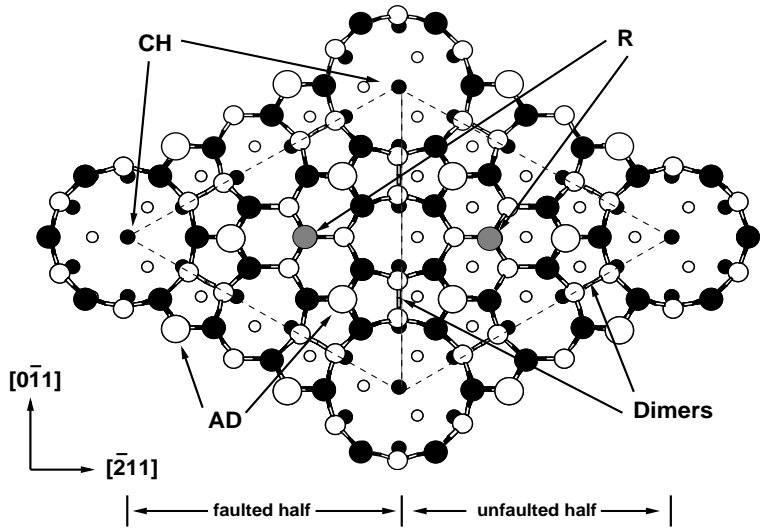}}\par}
\caption{\label{fig: fig1}{ Top view of Dimer-Adatom-Staking
fault structure of the \emph{Si(111)-5 $\times$}5 reconstruction. Circles
of decreasing diameter denote atoms in successive layers starting from the
top. Large white ones are adatoms, large black and grey ones belong to the
first layer with a stacking fault on the left half, smaller white ones
belong to the dimer layer, small black ones appear at the center of corner
holes and adjacent to the dimers, whereas remaining third layer atoms are
hidden under the adatoms, and small white ones belong to the fourth layer.
Adatoms(AD), rest atoms(R), corner holes(CH) atoms and the boundary of the
surface unit cell are indicated. The thin dotted-dashed line along the
short diagonal divides the faulted and unfaulted halves.}}
\end{figure}

The metastable 5$\times$5 reconstruction plays an important role in the
conversion from 2$\times$1 reconstruction, obtained upon cleavage, to the
stable 7$\times$7 reconstruction of the Si(111) surface \cite{feen1990}.
It exhibits the characteristic features of the DAS model first proposed
for Si(111)-7x7~\cite{7x7_1}, which are indicated in Fig.~\ref{fig: fig1}.
One important difference is that the 5$\times$5 reconstruction has the
same average density of atoms in surface layers as the bulk terminated
Si(111) surface, whereas additional Si adatoms must be supplied to form
the 7$\times$7 reconstruction.

\subsubsection{Computational Details}

We start with a ten-layers thick inversion-symmetric slab which has two
bulk truncated (111) surfaces at the top and bottom, encompassing two
adjacent 5$\times$5 surface unit cells. Periodic boundary conditions are
then maintained in the lateral directions. We have also repeated some of
the computations with a six-layers slab passivated by H atoms at the
bottom, similar to that used by Adams and Sankey~\cite{adams1992}.
Following these authors, the initial configuration on the free surface(s)
is set up by laterally shifting ten atoms in the surface layer on the left
side of each 5$\times$5 cell so that they are placed above fourth layer
atoms. Then five atoms at one edge of the other "unfaulted" triangular
half and the second-layer atom above the "corner hole" are placed at the
positions of the adatoms so that their bond lengths to the nearest
first-layer atoms are equal to the bulk interatomic distance $r_0$. During
the combined relaxation of electrons and ions, dimers spontaneously formed
along the boundaries between ``faulted'' and ``unfaulted'' halves shortly
after we started the damped molecular dynamics calculation. At the end, we
obtained the relaxed structure consistent with the DAS model illustrated
in Fig.\ref{fig: fig1}. Of the 25 dangling bonds per 5$\times$5 cell on
the truncated (111) surface only 9 are left: one on the corner hole atom
(CH), two on the restatom(R) and six on the adatoms(AD). The former three
are doubly occupied, whereas surface states derived from those on the
adatoms are partially occupied by the remaining three
electrons~\cite{adams1992}. Thus one expects the $Si(111)-5\times5$
surface have metallic properties just like $Si(111)-7\times7$. As a matter
of fact, {\it a posteriori} diagonalization revealed a tiny "energy gap"
($<
0.01 eV$) in our relaxed structure. One might also expect problems due to
the assumed double occupancy. In fact relaxation in a slab encompassing a
single $5\times5$ surface unit cell produced a distorted DAS structure
with twisted dimers. On the other hand, no such distortions have been
found in previous $O(N^3)$ computations of $Si(111)-7\times7$ with similar
slabs which relied on occupied eigenstates at the $\Gamma$-point(zero
wave-vector) obtained by diagonalization~\cite{qian1987} or
\cite{stich1992,brommer1992,kim1995-2} by total energy minimization {\it
with} orthonormality constraints~\cite{car-parrinello,payne1993}.
Deviations might arise owing to the small size of the localization
region(n=3) in our O(N) computations. Fortunately this is not the case, as
discussed in the following section, if slabs encompassing on even number
number of dangling bonds on each surface are used.

\subsubsection{Relaxed Geometry}

Our relaxed configuration exhibits common characteristic features found in
previous computations of the $5\times5$~\cite{adams1992,stich1992} and of
the $7\times7$~\cite{qian1987,brommer1992,kim1995-2} reconstructions. In
what follows we compare results obtained with Bowler's parametrization,
localization regions encompassing third neighbours and a $5\times10$
computational cell containing four free layers (not counting the adatoms)
and two fixed {\it Si} layers with the bottom one saturated by H atoms. In
order to check the influence of free and fixed(bulk-like) layers in the
slab, of passivation by H at the bottom, of the size of the localization
regions, and of the tight-binding parametrization we have performed many
calculations of the relaxed geometry of the $5\times5$ reconstruction.
From the result summarized in Appendix A one can conclude that the
influence of these factors on the final configuration is negligibly small.
This is a very encouraging result. Furthermore the bond lengths listed in
Table~\ref{bond1} are remarkably close (except for dimers) to those found
in recent TB calculations for the $7\times7$ reconstruction based on a
symmetric slab with one free layer less on each side~\cite{kim1995-2}.
Small deviations occur, however, compared to bond lengths extracted from a
LDA computation for the same system~\cite{brommer1992}. Similar deviations
also occur compared to an earlier TB computation for a H-passivated slab
with two less free layers~\cite{qian1987}. In that case, these deviations
might be due to that restriction or to the somewhat different
parametrization. More significant deviations appear when height
differences between adatoms and rest atoms in the two halves of the
surface unit cells are compared(see Table~\ref{table1}). Our computed
differences are larger than those found in previous computations for the
$7\times7$ reconstruction with a smaller number of free layers. On the
other hand, height difference between adatoms is only half of that found
in a pioneering computation for $Si(111)-5\times5$ based on a H-passivated
slab like ours~\cite{adams1992}. However, in contrast to that work, we
found that adatoms in each half are equivalent, just like corner hole and
central adatoms are on the $Si(111)-7\times7$ surface. These two
discrepancies are probably due to the {\it non self-consistent} LDA-based
approximation used in Ref.~\cite{adams1992}. TB computations, including
ours, are only rudimentary self-consistent if the Hubbard term is
included, but since the parameters are fitted to experimental
data~\cite{bowler1998} and/or self-consistent LDA
computations~\cite{kwon1994}, they can yield better results.
\begin{table}[!hbt]
\caption{\label{bond1}{ Computed average bond lengths(in \AA)
between Si atoms in surface layers of $Si(111)-5\times5$(present) and
$7\times7$ reconstructions. Here 1 - Adatom-first layer atom; 2 -
Adatom-second layer atom; 3 - Rest atom-second layer atom; 4 - Dimer; 5
- Dimer-third layer atom; 6 - Corner hole-fourth layer atom}}
\vskip 0.3cm 
{\centering \begin{tabular}{|c|l|c|c|c|}
\hline 
Bonds & present& Kim \emph{et al}\cite{kim1995-2}&
Qian \emph{et al}\cite{qian1987}& Brommer \emph{et al}\cite{brommer1992}\\
\hline 
1& 2.550& 2.58& 2.486& 2.49\\
\hline 
2& 2.575& 2.57& 2.471& 2.474\\
\hline 
3& 2.435& 2.43& 2.410& 2.376\\
\hline 
4& 2.417& 2.45& 2.463& 2.456\\
\hline 
5& 2.39& 2.39& 2.405& 2.396\\
\hline 
6& 2.408& & & 2.40\\
\hline 
\end{tabular}\par}
\end{table}

\begin{table}[!hbt] 
\caption{\label{table1}{ Computed height differences(in
\AA) between adatoms(AD) and rest atoms(R) in the faulted and unfaulted
halves of the $5\times5$ and $7\times7$ Si(111) surface unit cells. For
the $7\times7$ reconstruction, the two values given refer to the adatoms 
near corner holes and near the centers of each half cell.}} 
\vskip 0.3cm
{\centering 
\begin{tabular}{|c|c|c|c|c|} 
\hline Atoms & Present & Adams {\it et al}& Qian {\it et al} & Brommer
{\it et al}\\ 
\hline adatoms& 0.08 & 0.17 & 0.055, 0.07 & 0.047,0.030\\
\hline rest atoms& -0.05 & - & -0.01 & 0.03\\ 
\hline 
\end{tabular}\par}
\end{table}
\subsubsection{Convergence, Accuracy and Determination of Surface
Energies}

The accuracy, efficiency and performance of our O(N) TB computations can
be judged on the basis of the results reported in Table~\ref{local} for
the system described in the preceding section.

\begin{table}[!hbp]
\caption{\label{local} { Comparison of total energies
required computation times and numbers of iterations for a system of 300 
Si and 50 H atoms}} 
\vskip 0.3cm
{\centering \begin{tabular}{|c|c|c|c|c|}
\hline 
Relaxation& Quantity& 2 LR& 3 LR& Diagonalization\\
\hline 
Initial& Total Energy(\emph{eV})& -12896.83& -12918.38& -12932.55\\
\cline{2-2} \cline{3-3} \cline{4-4} \cline{5-5} 
\multicolumn{1}{|c|}{(Electrons)}& Computation time(hrs.)& 4.5&
23 &
\\
\cline{2-2} \cline{3-3} \cline{4-4} \cline{5-5} 
\multicolumn{1}{|c|}{}&
CG steps&
3400 &
5500&
\\
\hline 
\multicolumn{1}{|c|}{Electrons }&
Total Energy(\emph{eV})&
-12958.05 &
-12977.68&
-12987.32\\
\cline{2-2} \cline{3-3} \cline{4-4} \cline{5-5} 
\multicolumn{1}{|c|}{and Atoms}&
Computation time(hrs.)&
6.5&
22 &
\\
\cline{2-2} \cline{3-3} \cline{4-4} \cline{5-5} 
\multicolumn{1}{|c|}{}&
MD steps&
1848 &
1778 &
\\
\hline 
\end{tabular}\par}
\end{table}

Computations were performed on a single processor of a DEC-Alpha 8400
machine; nLR refers to unconstrained minimizations with localization
regions encompassing neighbours connected by n bonds. Diagonalization was
performed at the end of the 3LR minimization. The first three rows refer
to the unbiased but costly minimizations starting with orthonormalized
LOs with random coefficients for the initial configuration. It is
gratifying that the subsequent combined optimization of LO coefficients
and atomic positions takes about the same time. More importantly the ratio
of CG to MD steps implies that only three CG steps per MD step are on the
average required to reconverge the coefficients.

A comparison of the last two columns suggests that 3LR minimization yields
total energies with a relative accuracy of about $10^{-4}$. The specified
tolerance on the relative energy difference between successive CG
iterations(our convergence criterion) was, of course, much smaller. Note
that diagonalization yields eigenstates corresponding to the
$\Gamma$-point of the computational supercell. More accurate total
energies could be obtained by including occupied eigenstates with nonzero
parallel wave-vectors in Eq.(\ref{band}) or by increasing the lateral
dimensions of the supercell(the only alternative in the case of O(N)
computations). The surface energy differences $\Delta E_ s$ which determine
the relative stability of possible reconstructions are usually
approximated by differences between total energies per projected
$1\times1$ surface unit cell computed in the same computational slab with
the desired reconstruction on one face and same reference structure on the
other. This approximation is reasonable if the system is sufficiently
large, in particular to effectively decouple the two faces. On the other
hand, an upper bound on the error in $\Delta E_ s$ is given by the product
of the number of layers times the energy per unit cell of the substrate
times the relative accuracy. According to Table~\ref{local} this amounts
to $\approx 0.05 eV$. Computed values of $\Delta E_ s$ at the 3LR level
with respect to the truncated Si(111) surface are only -0.19 eV and -0.13
eV for the symmetric and H-passivated slabs described in section 3.1.1.
The difference between those two values is disturbingly close to our
estimated error bound. Furthermore, previous estimates of surface energy
differences between the $5\times5$ and the $2\times1$ and $7\times7$
reconstructions, which are relevant for understanding their
growth~\cite{feen1990}, from self-consistent LDA computations amount to
-0.06 eV~\cite{brommer1992} and 0.02 eV~\cite{stich1992}, respectively.
This implies that O(N) computations beyond the 3LR level of accuracy will
be required to distinguish them reliably. On the positive side, note that
$\Delta E_ s$ = -0.15 eV is found upon diagonalization for both 
above-mentioned slabs. Finally, the rather different values of $\Delta E_ s$
obtained in Refs.~\cite{qian1987} and \cite{adams1992}, namely -0.395 eV
and 0.56 eV, suggest that TB parameterization and non-selfconsistency are
delicate issues which should be addressed.      

\subsection{Si(001)-c(4x2) Reconstruction}

A truncated 1$\times$1 Si(001) surface contains many unsaturated bonds and
the system tends to minimize its energy by reconstructing its surface.
Theoretical\cite{chadi1979,khan1989,yin1981} and
experimental\cite{tromp1985} evidence shows that this reconstruction
causes dimers to appear on the surface; i.e. surface atoms move toward
each other to form pairs. Furthermore, these dimers are tilted and
asymmetric with respect to terminated bulk. The
dimers can arrange themselves in various patterns on the surface and thus
many reconstructions of the surface are possible. 

\begin{figure}[!hbt]
{\centering 
\begin{tabular}{cc}
\subfigure[ideal surface]{\resizebox*{6cm}{4cm}
{\includegraphics{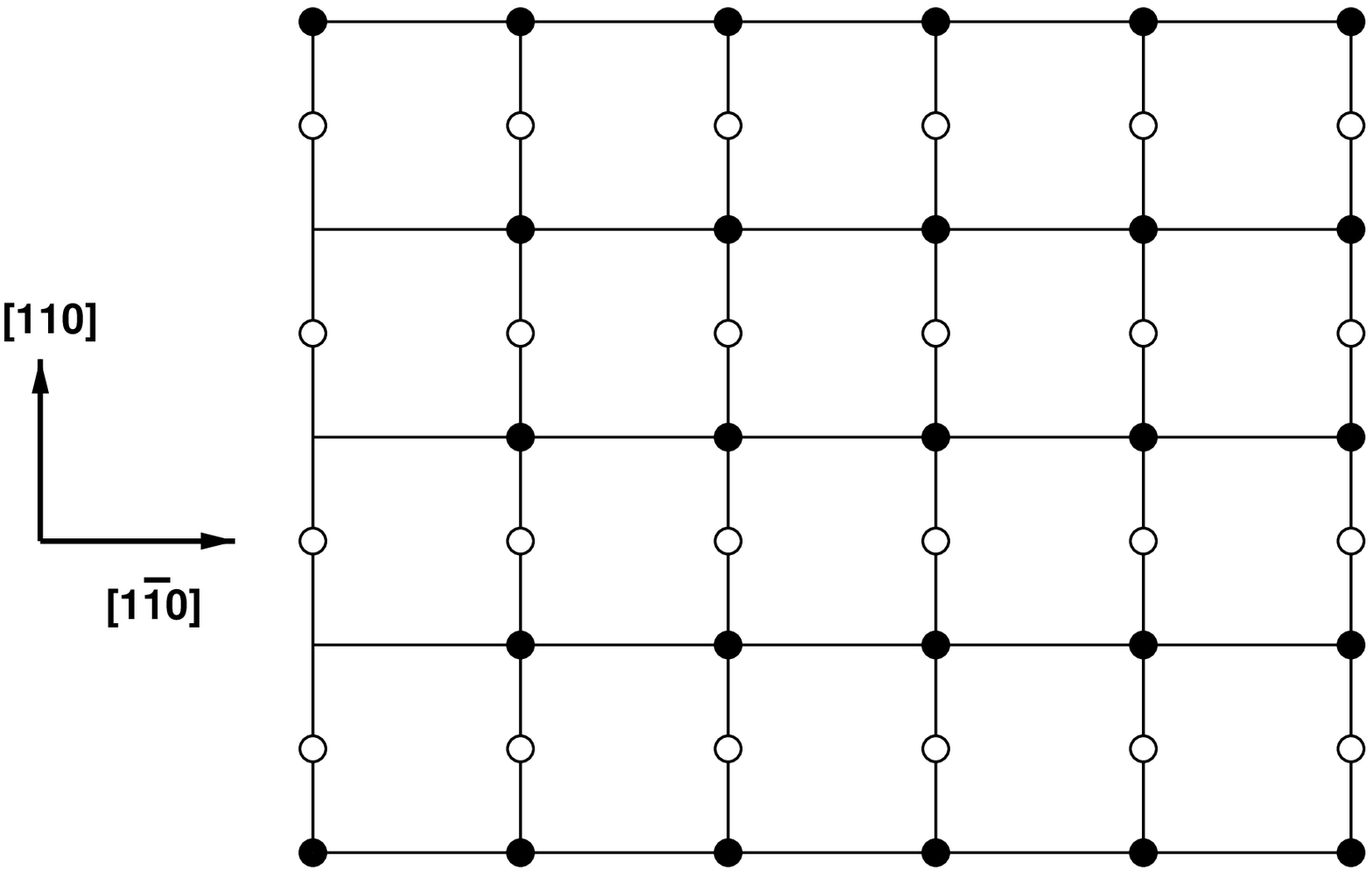}}} ~~~&
~~~\subfigure[reconstructed surface]{\resizebox*{5cm}{4cm}
{\includegraphics{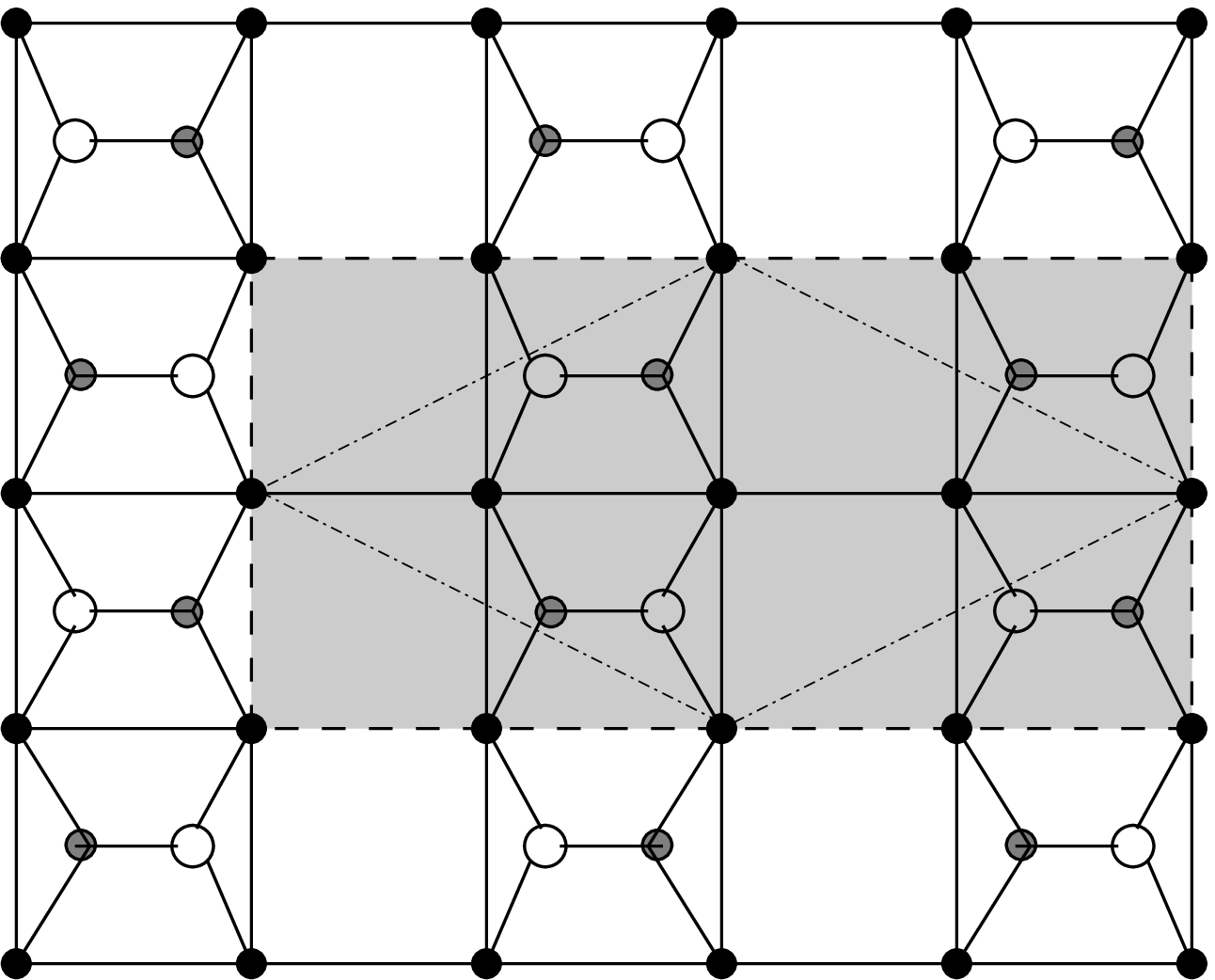}}} \\
\end{tabular}\par}
\caption{\label{si001top}{ Top views of the unreconstructed
(bulk truncated) \emph{Si(001)} surface and of its c(4$\times$2)
reconstruction. The shadowed area delimits the surface unit cell used in
our simulation. Black circles denote second layer atoms, grey and white
circles denote dimerized surface atoms. The white atoms lie higher than
the grey counterparts. The primitive c(4$\times$2) surface unit cell is
outlined by the
dot-dash line in (b).}}
\end{figure}

We have made a less extensive set of O(N) computations for the Si(001)
surface. Following a period of controversy, several LDA studies,
in particular the exhaustive one of Ramstad {\it et al}~\cite{ramstad1995}
have confirmed that the $Si(001)-c(4\times2)$ reconstructions
schematically illustrated in~Fig.\ref{si001top}(b), and first predicted in
a pioneering TB computation~\cite{chadi1979}, has in fact the lowest
energy.
The $p(2\times2)$ structure, characteristic by identical [110] rows with
alternatively tilted Si-dimers, is only marginally metastable, while the
2$\times$1 structure with untilted, symmetric dimers is unstable at low
temperatures and 0.1 eV higher in energy per projected (1$\times$1) 
surface unit cell~\cite{ramstad1995}.

We report O(N) computations performed with Bowler {\it et
al}~\cite{bowler1998}'s parameterization, n=3 LRs in the 4$\times2$
surface unit cell indicated by shading in Fig.\ref{si001top}(b). Our
computational slab consisted of 6 free Si layers and two fixed layers,
with the pairs of dangling bonds at the bottom passivated by H atoms in
order to approximate a connection to the bulk crystalline silicon
substrate. Starting from a configurations with slightly preformed untilted
dimers, our computations converged towards a $c(4\times2)$ reconstruction,
although the $p(2\times2)$ was obtained in some cases.

\begin{table}[htp]
\caption{\label{si_c4x2}{ Computed atom displacements from
ideal truncated bulk positions (in \AA)}}
\vskip 0.3cm
{\centering
\begin{tabular}{|c|c|c|c||c|c|c||c|c|c||c|c|c|} 
\hline &
\multicolumn{6}{|c||}{ Tight-binding calculations}&
\multicolumn{6}{|c|}{LDA calculations}\\ 
\cline{8-13} 
\cline{2-4}
\cline{5-7} 
\cline{8-10} 
\cline{11-13} 
\multicolumn{1}{|c|}{Layer}&
\multicolumn{3}{|c||}{ present}&
\multicolumn{3}{|c||}{Chadi\cite{chadi1979} }&
\multicolumn{3}{|c||}{Ramstad \emph{et al}\cite{ramstad1995}}&
\multicolumn{3}{|c|}{Northrup\cite{northrup1993}}\\ 
\cline{8-10}
\cline{11-13} 
\cline{2-2} 
\cline{3-3} 
\cline{4-4} 
\cline{5-5} 
\cline{6-6}
\cline{7-7} 
\cline{8-8} 
\cline{9-9} 
\cline{10-10} 
\cline{11-11}
\cline{12-12} 
\cline{13-13} 
\multicolumn{1}{|c|}{}& $\Delta x$ & $\Delta y$ & $\Delta z$ & $\Delta x$ 
& $\Delta y$ & $\Delta z $ & $ \Delta x $ & $ \Delta y $& $ \Delta z $& 
$ \Delta x $& $ \Delta y $& $ \Delta z $\\
\cline{2-2} \cline{3-3} \cline{4-4} \cline{5-5} \cline{6-6} \cline{7-7}
\cline{8-8} \cline{9-9} \cline{10-10} \cline{11-11} \cline{12-12}
\cline{13-13} 
\hline 
1& 0& 0.629& 0.045& 0.& 0.46& 0.04& 0& 0.68& -0.05&
0& 0.61& -0.04\\ 
\hline 
1& 0& -0.821& -0.383& 0.& -1.08& -0.435& 0& -0.95&
-0.788& 0& -1.04& -0.74\\ 
\hline 
2& 0.09& 0.115& -0.011& 0.& 0.115& -0.014&
0.117& 0.12& -0.086& 0.008& 0.1& -0.07\\ 
\hline 
2& -0.09& -0.115& -0.01& 0.& -0.115& -0.014& -0.12& -0.108& -0.079&
-0.008&
-0.1& -0.07\\ 
\hline 
3& 0.& 0.& -0.2& 0.& 0& -0.12& 0& 0.009& -0.223& 0& 0& -0.19\\ 
\hline 
3& 0.008& 0.& 0.117& 0.& 0& 0.11& 0.02& 0.003& 0.066& 0.001& 0& 0.05\\
\hline
4& 0& 0.023& -0.132& 0.& 0& -0.07& 0& 0.006& -0.154& 0& 0& -0.12\\ 
\hline
4& 0& 0.& 0.08& 0.& 0& 0.07& 0& 0.005& 0.069& 0& 0& 0.04\\ 
\hline 
5& 0& 0.049& -0.01& 0.& 0.034& 0.& 0& 0.043& -0.03& & & \\ 
\hline 5& 0& -0.067& -0.31& 0.& -0.034& 0.& 0& -0.038& -0.03& & & \\ 
\hline
\end{tabular}\par}
\end{table}
As can be seen from Table~\ref{si_c4x2}, the resulting pattern of atomic
displacements is reproduced correctly, the relaxed coordinates being
within the spread of values from previous computations. Relevant
deviations are more evident in Table~\ref{si(001)}. The dimer tilt, being
energetically easy, is quite sensitive to the level of approximations. The
deviation of the dimer bond length from the bulk Si-Si distance(2.35\AA),
in the opposite direction to LDA prediction is more serious and
disappointing. Indeed, Bowler {\it et al}~\cite{bowler1998} claimed that
their parameterization would cure this discrepancy. More computations are
needed to check the extent to which the above-mentioned deviations are
affected by computational approximations, and to extract reliable surface
energy differences.
\begin{table}[!htb]
\caption{\label{si(001)}{ Computed bond lengths and tilt
angle of surface dimer.}}
\vskip 0.3cm
{\centering \begin{tabular}{|c|c|c|c|}
\hline 
Bonds& present& Ramstad \emph{et al}\cite{ramstad1995}& 
Chadi\cite{chadi1979}\\
\hline 
I& 2.425& 2.29& 2.35\\
\hline 
II& 2.345& 2.31& 2.332\\
\hline 
III& 2.367& 2.35& 2.390\\
\hline 
IV& 2.379& 2.37& 2.398\\
\hline 
V& 2.377& 2.35& 2.34\\
\hline 
$\alpha$& 10.2$ ^{o} $& 18.8$ ^{o} $& 11.68$ ^{o} $\\
\hline 
\end{tabular}\par}
\vspace{0.1cm}
{\centering \subfigure{\resizebox*{7.5cm}{3.75cm}
{\includegraphics{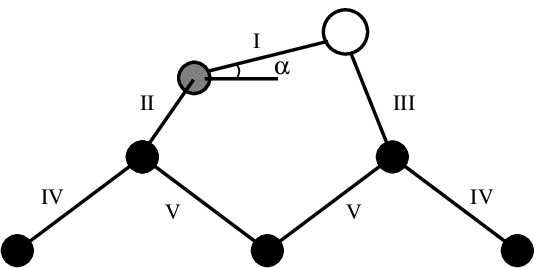}}}\\
{\centering Inset: Side view of tilted dimer and of second and third
layer atoms}\\
\par}
\end{table}

{\it A posteriori} diagonalization gives a 0.9 \emph{eV} band gap, a value
which is reasonable~\cite{northrup1993}, but should not be taken too
seriously because the TB parameters are fitted to occupied valence bands
and to ground-state properties. The existence of a band gap is consistent
with the known semiconducting nature of the $c(4\times2)$.

\section{Conclusions}

Summarizing the results described in section 3 and Appendix A, we conclude
that the local-orbital-based linear scaling TB scheme proposed by Kim {\it
et al}~\cite{kim1995} reproduces the correct configurations of
representative silicon surface reconstructions. Except for a few
discrepancies which can be traced to inadequencies of the tight-binding
description itself, satisfactory geometries are obtained with the simpler
parametrization of Bowler {\it et al}~\cite{bowler1998} and with local
orbitals constrained to vanish beyond second nearest neighbours. This even
applies to the metallic $Si(111)-5\times5$ surface provided that the
computational surface unit cell encompasses on even number of dangling
bonds. These encouraging results open the door to applications to larger
systems exploiting the linear-scaling capability of this new
computational scheme, e.g. involving interactions on and between silicon
clusters and surfaces exposing different faces with or without passivating
hydrogen atoms. 

It is, however, important to keep in mind that a higher level of accuracy
appears required to quantitatively describe surface energy differences
between alternative (meta)stable structures. This aspect is currently
under study.

\section{Acknowledgements}

This work was supported by the Swiss National Foundation for Scientific
Research under the programm NFP36 "Nanosciences". The first two authors
are grateful for the facilities and services provided by the computing
center and the Institute of Physics of University of Basel. They wish to
thank Prof. H.-J. G\"untherodt for his encouragement, D. Bowler and R.
H\"arle for discussions.


\appendix

\section{Influence of Various Factors}

As explained in section 2.3, the use of local orbitals assumed to
vanish outside finite localization regions is main approximation leading
to linear scaling. O(N) TB computations are efficient if the range of TB
interactions and size of the LR(the number of neighbours connected to the
central atom by n bonds) can be chosen as small as possible without unduly
sacrificing accuracy. For this reason we have performed test calculations
with n=2 and n=3 LRs for the two TB parametrizations described in section
3.1 and different slabs. Representative bond lengths obtained for the
$Si(111)-5\times5$ reconstruction are shown in Table~\ref{bond}. Column 1
and 2 show the results obtained with the complex and simpler
parametrizations using a symmetric slab with 8 layers(not counting
adatoms) and n=3 LRs. Column 3 shows the results obtained with the same LR
and using TB parameters as in column 2 but for a slab with 6 layers,
passivated by hydrogen atoms at the bottom. Column 4 shows results
obtained with for the same TB parameters and slab as in column 3, but
using n=2 LRs. From the results one can see that these factors have
little influence on the final relaxed geometry. 
\begin{table}[hbp]
\caption{\label{bond}{ 
Comparison of bond lengths in surface layers of the $Si(111)-5\times5$
reconstruction computed for different TB parametrizations, slab and
localization regions. Here 
1 - Adatom-first layer atom; 
2 - Adatom-second layer atom; 
3 - Rest atom-second layer atom; 
4 - Dimer; 5 - Dimer-third layer atom; 
6 - Corner hole-fourth layer atom}}
\vskip 0.3cm
{\centering \begin{tabular}{|c|c|c|l|c|}
\hline 
\multicolumn{1}{|c|}{}&
\multicolumn{2}{|c|}{parametrization (symmetric slab, 3
LR)}&
\multicolumn{2}{c|}{LR(with H, Bowler
\emph{et al}'s)}\\
\hline 
Bonds& Kwon \emph{et al}~\cite{kwon1994}'s& Bowler
\emph{et
al}~\cite{bowler1998}'s& 3 LR& 2 LR\\
\hline 
1& 2.545& 2.56& 2.550& 2.545\\
\hline 
2& 2.58& 2.58& 2.575& 2.580\\
\hline 
3& 2.44& 2.44& 2.435& 2.446\\
\hline 
4& 2.41& 2.42& 2.417& 2.414\\
\hline 
5& 2.39& 2.39& 2.39& 2.39\\
\hline 
6& 2.44& 2.41& 2.408& 2.410\\
\hline 
\end{tabular}\par}
\end{table}
\section{The matrix element $<\alpha _{Jl}|\hat{H}_{TB}|\alpha _{J^{'}l^{'}}>$}
When $ J=J^{'} $, the matrix is given by
\[<\alpha _{Jl}|\hat{H}_{TB}|\alpha _{J^{'}l^{'}}>=\left\{ \begin{array}{ll}
E_{s} & \begin{array}{ll}
if & l=l^{'}=s
\end{array}\\
E_{p} & \begin{array}{ll}
if & l=l^{'}=p_{x},p_{y},p_{z}
\end{array}\\
0 & \begin{array}{ll}
if & l\neq l^{'}
\end{array}
\end{array}\right. \]


When $J\neq J^{'}$, the various matrix elements can be written as,

\begin{eqnarray}
E_{Js,J^{'}s}&=&V_{ss\sigma } \\ \nonumber
E_{Js,J^{'}p_x}&=&-E_{Jp_x,J^{'}s} \;=\; lV_{sp\sigma }\\  \nonumber
E_{Js,J^{'}p_y}&=&-E_{Jp_y,J^{'}s} \;=\; mV_{sp\sigma }\\  \nonumber
E_{Js,J^{'}p_z}&=&-E_{Jp_z,J^{'}s} \; = \; nV_{sp\sigma}\\  \nonumber
E_{Jp_{x},J^{'}p_{y}}&=&-E_{Jp_{y},J^{'}p_{x}} \; =\;lm(V_{pp\sigma
}-V_{pp\pi })\\  \nonumber
E_{Jp_{x},J^{'}p_{z}}&=&-E_{Jp_{z},J^{'}p_{x}} \;= \; ln(V_{pp\sigma
}-V_{pp\pi })\\  \nonumber
E_{Jp_{y},J^{'}p_{z}}&=&-E_{Jp_{z},J^{'}p_{y}} \;= \; mn(V_{pp\sigma
}-V_{pp\pi })\\  \nonumber
E_{Jp_{x},J^{'}p_{x}}&=&E_{Jp_{x},J^{'}p_{x}} \;= \; l^{2}V_{pp\sigma
}+(1-l^{2})V_{pp\pi }\\  \nonumber
E_{Jp_{y},J^{'}p_{y}}&=&E_{Jp_{y},J^{'}p_{y}} \; = \; m^{2}V_{pp\sigma
}+(1-m^{2})V_{pp\pi }\\  \nonumber
E_{Jp_{z},J^{'}p_{z}}&=&E_{Jp_{z},J^{'}p_{z}} \; = \; n^{2}V_{pp\sigma
}+(1-n^{2})V_{pp\pi }
\end{eqnarray}
here $l,m,n$ are the direction cosines of the vector 
$\mathbf R_{J^{'}}-\mathbf R_J$:
\[ l=\frac{R_{J^{'}x}-R_{Jx}}{|\mathbf R_{J^{'}}-\mathbf R_{J}|}\]
\[ m=\frac{R_{J^{'}y}-R_{Jy}}{|\mathbf R_{J^{'}}-\mathbf R_{J}|}\]
\[ n=\frac{R_{J^{'}z}-R_{Jz}}{|\mathbf R_{J^{'}}-\mathbf R_{J}|}\]

\end{document}